# Particle Mass Oscillation through Tachyon Interaction

L. Nanni†
University of Ferrara
44100-Ferrara, Italy
†E-mail: luca.nanni68@gmail.com

**Abstract**: In this study, a novel theory to investigate the mass oscillation of particles is proposed. It has been proven that, at high-energy conditions, the fermion field described by Dirac's Lagrangian interacts with the half-integer spin tachyon field with negative energy, causing the formation of composite particles whose mass depends on the total angular momentum. The proposed theory is based on a new interpretation of the Majorana equation for particles with arbitrary spin and shows that mass oscillation is a phenomenon in which the component of particle decay prevails over that of mixing mass states. Using the kinematic of Lemke for spacelike particle decay, we propose a mechanism able to explain the neutrino flavour change. The proposed mechanism is also investigated concerning the shape of its spectrum. Finally, the Lagrangian field of composite particles is formulated.

1. Introduction

The mass oscillation of particles is one of the unsolved problems of particle physics and, together with those concerning the origin of masses, flavours and quantum numbers, make the current Standard Model an unsatisfactory and incomplete theory [1-2]. To date, the only case of mass oscillation experimentally observed is the flavour change of the neutrino during free flight [3-4], a phenomenon for which there is not yet a robust theory capable of explaining it. To overcome this impasse, the route of supersymmetry theory has been undertaken, which is expected to be the best candidate to definitively resolve all the current

*shortcomings* of the Standard Model [5-7]. But, in a scenario still consisting of uncertainties [8], it is considered worth investigating other *niche* theories, some of which are perhaps still too speculative [9-12], which could help to end this impasse and facilitate the formulation of a more solid and complete quantum particle theory. To this purpose, the introduction of the tachyon field, which is able to interact with ordinary matter [13], could help to explain the mass oscillation of particles [14]. For fermions, this theory has already been partly developed [15], and the aim of this study is to complete it. The theory developed in [15], based on a new interpretation of the Majorana theory for particles with arbitrary spin [16], shows that a massive free particle with half-integer spin at high energies can interact with a negative energy tachyon field forming a composite particle whose real mass is lower than the initial one. This result can be extended to antiparticles that will interact with positive energy tachyonic fields. A similar result has already been found by another author [17], but, in our case, the mass spectrum is quantized, and this facilitates the formulation of equations able to explain the *transmutation* of a fermion in one of the possible flavours or the mass oscillations of hadronic particles and atomic nuclei [14].

Using the results obtained in [15] and [18], we get the resonance frequency and the probability of coupling between fermionic and tachyonic fields, showing that their interaction stabilizes the high energy states of the particle by decreasing its real mass and that this stabilization has a limit beyond which the coupling becomes unfavourable. The obtained equation describing mass oscillation is then compared with that obtained assuming a mechanism due exclusively to the mixing phenomenon between two possible mass states. From this comparison, it is shown that the decay contribution is preponderant compared to that of mixing and increases with the increase of the coupling with the tachyon field. Based on this result, a mechanism of decay mediated by a vector boson, similar to beta decay, which allows estimating the range and the strength of the interaction force, is proposed. This mechanism is congruent with Lemke's kinematics [19], which describes decays of ordinary particles with tachyon emission.

Finally, the Lagrangian field for the bradyon-tachyon composite particle (Majorana field) is formulated. This leads to a theory that can further extend the Standard Model (the starting point is the Majorana equation which has been formulated to be invariant to the Lorentz group in its infinite-dimensional representation) to make it capable of explaining the mass oscillation of light particles and of predicting the existence of an infinite number of mass oscillations that potentially characterize the current particle scenario.

## 2. Bradyon-Tachyon Composite Particles

The energy-momentum relation for a composite bradyon-tachyon particle, whose field is given by the interaction between the fermion field (with positive energy) and the tachyonic one (with negative energy), is [15]:

$$E^2(J) - E^2(J_0) = [\boldsymbol{p}^2(J) - \boldsymbol{p}^2(J_0)]c^2 - \frac{\left(J+\frac{1}{2}\right)^2 - 1}{\left(J+\frac{1}{2}\right)^2} m_0^2 c^4, \tag{1}$$

where $J_0$ is the spin of the fermion in the centre-of-mass frame (in our case, $J_0 = 1/2$). Considering that $\boldsymbol{p}^2(J_0) = 0$ and that $E^2(J_0) = m_0^2 c^4$, Eq. (1) may be rewritten as:

$$E^2(J) = \boldsymbol{p}^2(J)c^2 - m_0^2 c^4, \tag{2}$$

proving that, in the centre-of-mass frame, all the components of the Majorana field are tachyonic. This means that, when the relative velocity of the particle is zero, the fermionic field is completely decoupled from the tachyonic one. In this situation, the free tachyon field has an imaginary continuous mass spectrum, as predicted by the Majorana equation [16]. But, as soon as the reference frame is changed, the fermion velocity becomes different from zero and vibrations with $J > 1/2$ are *activated*, leading to the coupling of fermionic field with the tachyonic one. These *resonance* frequencies give rise to the bradyonic mass spectrum [16]:

$$m(J) = \frac{m_0}{(J + 1/2)}. \tag{3}$$

This coupling quantizes the tachyon field, producing an imaginary mass spectrum that is no longer continuous but discrete:

$$\mu(J) = i \frac{\sqrt{\left(J + \frac{1}{2}\right) - 1}}{\left(J + \frac{1}{2}\right)} m_0 \quad J = n/2, \quad n \in N. \tag{4}$$

Therefore, the Majorana field for the composite particles has a real mass spectrum given by:

$$m^2(J) = m_0^2 + \mu^2(J) = \frac{m_0^2}{\left(J + \frac{1}{2}\right)^2}. \tag{5}$$

The Lagrangian density of the tachyon field with negative energy is [15]:

$$\mathcal{L}_{L_-} = \sum_J \frac{g(\varepsilon(J))}{\sqrt{\varepsilon(J)}} \left[ \Lambda_J^{-1} \bar{\psi}_{D_+} (i\gamma^\mu \partial_\mu + im) \Lambda_J^{-1} \psi_{D_+} \right], \tag{6}$$

where $\psi_{D_+}$ is the Dirac field with positive energy, $\varepsilon(J) = \sqrt{\left(J + \frac{1}{2}\right) - 1} / \left(J + \frac{1}{2}\right)$ and $\Lambda_J^{-1}$ is the superluminal Lorentz transformation (SLT) matrix, depending on quantum number $J$. The probability that the fermionic field of mass $m_0$ interacts with the tachyonic one is given by [18]:

$$P(J) = \sqrt{\beta^n - \beta^{n+1}} \quad n = J - \frac{1}{2}, \quad \beta = \frac{u}{c}, \tag{7}$$

where $u$ is the particle velocity. For a given value of $J$, there exists a particle velocity which makes the probability maximum:

$$\frac{\partial P(J)}{\partial \beta} = 0 \Rightarrow \beta_{max} = \frac{n}{n+1} = \frac{J - 1/2}{J + 1/2}, \tag{8}$$

to which corresponds the following Lorentz factor:

$$\gamma_{max}(J) = \frac{J + 1/2}{\sqrt{2J}}. \tag{9}$$

Therefore, the fermion energy that makes maximum the coupling with the tachyon filed is:

$$E_{max}(J) = \frac{J + 1/2}{\sqrt{2J}} m_0 c^2, \tag{10}$$

to which corresponds the angular *resonance* frequency:

$$\omega_{res}(J) = \frac{J + 1/2}{\hbar\sqrt{2J}} m_0 c^2. \tag{11}$$

Once set, the particle velocity for each value of $J$ in Eq. (10) gives the *resonance* frequency at which the fermion field couples with the tachyonic one. Probability $P(J)$ tends asymptotically to zero increasing $J$, i.e. the initial fermion is less and less coupled with the negative energy tachyon field, proving that the formation of high-$J$ composite systems is disadvantaged. This result is confirmed by the fact that Eq. (8) holds for $J = 1/2$, i.e. the fermion tends to remain in the state with the lowest $J$ without interacting with the tachyon field. However, the probability that the coupling occurs is never zero, and, if it takes place (at the *resonance* frequency), then the fermion mass decreases according to Eq. (3) and the composite particle accelerates. Increasing the velocity, the particle becomes unstable as its energy progressively increases; however, when it is coupled to the negative energy tachyon field, the particle that is formed will have a lower energy because of the mass decreasing. This reasoning also explains why particles with high mass tend to interact less with the tachyon field: to be stabilized they must be coupled with a high negative energy tachyon, which is very unstable (to have a highly energetic tachyon, it is necessary that its velocity approaches that of light, i.e. in a particularly unfavourable kinematic region for a tachyon). It is concluded that the particles most likely to interact with the negative energy tachyon field are those with a very small mass.

Before going further, it must be clarified that the composite particle is given by the interaction of the initial fermion and tachyon pairs with parallel spin, so as to obtain values of $J$ equal to $n/2$.

As a result of coupling with the tachyon field, the composite particle increases its velocity and, consequently, also its probability to interact again with a new pair of tachyons. After each coupling, the value of the quantum number $J$ increases by one unit. Therefore, the coupling constant between the two fields is directly proportional to the particle velocity and inversely proportional to $J$. Moreover, having proved that the particle tends to remain in a low $J$-state, the bradyon-tachyon interaction force is certainly of a weak nature.

3. **Bradyon-Tachyon Interaction**

Since the coupling of the Dirac field with the negative energy tachyon field has the *effect* of decreasing the real mass, then the interaction force acts at long range. At the limit $u \to c$, the mass particle tends to zero and $J$ tends to infinity. Using the Lemke theory on the kinematic of decaying of an ordinary particle [19] and supposing that the lepton number is conserved, the particle may decay in a bradyon, tachyons and luxons. In particular, if $\boldsymbol{q}_i^2 \geq M^2$ ($\boldsymbol{q}_i$ is the four-impulse of the emitted fermion and luxons, while $M$ is the mass of the decaying particle), then not all tachyons may have positive energy, while, if $\boldsymbol{p}_i^2 \geq M^2$ ($\boldsymbol{p}_i$ is the tachyon four-impulse), then only one tachyon is emitted with negative energy. In the theory we are formulating, if the lepton number is conserved, then the following mechanism holds:

$$m(J+1) \to m(J) + \bar{t}(\mu) + t(\mu') + l_x. \qquad (12)$$

In other words, the conservation of the lepton number implies the production of an antitachyon (with negative energy) and a tachyon (with positive energy) having a parallel half-integer spin. The imaginary masses $\mu$ and $\mu'$ do not necessarily have to be equals; what is required is that the sum of their energy is negative:

$$E(\mu) + E(\mu') < 0. \qquad (13)$$

Eq. (13) complies with the results obtained applying the Lemke kinematics [19] to the Majorana theory [20], which predicts the most probable decaying mechanism leading to the emission of bradyons and tachyons with negative energies. The production of only bradyons occurs if Eq. (13) is equal to zero. Finally, considering the requirement of total spin conservation, we have:

$$\begin{cases} (J+1) = J + 1/2(\bar{t}_\mu) + 1/2(t_{\mu'}) + 0(l_x) \\ (J+1) = J \pm 1/2(\bar{t}_\mu) \mp 1/2(t_{\mu'}) + 1(l_x) \end{cases}. \tag{14}$$

In the first case, the tachyon and antitachyon have parallel half-integer spin, and, consequently, the emitted luxon is of the scalar type. In the second case, the tachyon and antitachyon have antiparallel spin, and the emitted luxon is of the vector type with spin 1. The possible scalar luxon emitted is a Goldstone boson, which is formed in systems with spontaneous symmetry violation [21]. The formation of a Goldstone particle holds with the Majorana theory because it has an infinite degree of freedom [22]. In fact, the Majorana theory has been formulated using the infinitesimal transformations of the Lorentz group, which are the starting point for continuous symmetry, a necessary condition to have a Goldstone boson.

We have seen that the probability of coupling between the fermion and tachyon fields is considerable when the particle velocity approaches the speed of light. In the theory we are formulating, neither the range nor the intensity of the interaction between the two fields is known. Invoking the uncertainty principle in the form $\partial E \partial t \geq \hbar$ and denoting the interaction range by $r_0$, we have:

$$\partial t \geq \frac{r_0}{c} \; ; \; \partial E \geq m(\nabla J)c^2 \; ; \; m(\Delta J) = \frac{m_0}{(n+2)(n+2)}. \tag{15}$$

In other words, we suppose that the interaction time is greater than $r_0/c$. Therefore, the interaction range is:

$$r_0 \cong \frac{\hbar}{m(\Delta J)c} = \frac{\hbar}{m_0 c}(n+2)(n+2) = \lambda_{Compton}(n+2)(n+2). \tag{16}$$

Eq. (16) shows that the interaction range becomes greater as the mass of the fermion in the centre of the mass frame becomes smaller and the quantum number $J$ becomes greater. In this scenario, the neutrino is the fermion that most interacts with tachyon field because of its very small mass and its velocity being very close to the speed of light. Considering that electron neutrino mass is $4.25 \cdot 10^6$ times lower than electron one [23] and that its velocity is at least 99.99% that of light [24] (to which corresponds the value $n = 9.9 \cdot 10^3$) and using Eq. (16), the calculated interaction range is of the order of Km, i.e. a value comparable with the travel distance at which neutrino flavour oscillation is observed [25]. Therefore, the theory we are formulating could be the gateway to explain the flavour change of neutrinos during their free flight.

The heavier particles, on the other hand, interact with the tachyon field only at shorter distances, which are less than or equal to the sub-atomic ones. However, experience seems to show no mass oscillation for massive particles other than neutrinos, even if we consider nuclear distances. This suggests that the interaction between ordinary and tachyonic matter is of a weak type and is completely *covered* by other more intense interactions that take place at nuclear or sub-nuclear distances.

## 4.  Decay Mechanism

In this section, we propose a decay mechanism for the composite particles obtained by bradyon-tachyon interaction. Considering that bradyons and tachyons are stabilized by decreasing and increasing their velocities, respectively, and that their interaction is of a weak type, the decay of the composite particle with $J > 1/2$ could follow a mechanism very similar to that of beta decay, with a new mediator boson that we denote by $N$:

$$m(J) \overset{N}{\Rightarrow} m(J-1) + \bar{t}_\mu + t_{\mu'} + \gamma. \tag{17}$$

The proposed mechanism holds if the mediator boson has spin 1 and if it satisfies the following constraint:

$$\mu(\bar{t}_\mu) + \mu(t_{\mu'}) = \mu(J) = i\frac{\sqrt{\left(J+\frac{1}{2}\right)-1}}{\left(J+\frac{1}{2}\right)}m_0. \tag{18}$$

The possible luxon ($\gamma$), as already mentioned in the previous section, is a scalar Goldstone boson. To calculate the matrix $\Lambda$ which transforms the spinor $m(J)$ into $m(J-1)$, let us consider the case where $m(J-1) = m_0$. The matrix must satisfy the following equation for each Dirac spinor:

$$\Lambda_J \psi_J = \psi_0. \tag{19}$$

The up and down spinors with positive and negative energy for a given Dirac fermion with mass $m_J$ are the columns of the following matrix [26]:

$$\sqrt{E(J)+m(J)}\begin{pmatrix} \mathbb{1} & & \frac{p_z(J)}{E(J)+m(J)} & \frac{p_-(J)}{E(J)+m(J)} \\ & & \frac{p_+(J)}{E(J)+m(J)} & -\frac{p_z(J)}{E(J)+m(J)} \\ \frac{p_z(J)}{E(J)+m(J)} & \frac{p_-(J)}{E(J)+m(J)} & & \mathbb{1} \\ \frac{p_+(J)}{E(J)+m(J)} & -\frac{p_z(J)}{E(J)+m(J)} & & \end{pmatrix} \tag{20}$$

where $p_\pm = p_x \pm ip_y$. Masses, energies and impulses of particles with a different quantum number $J$ are linked by the following relationships:

$$\begin{cases} m(J) = \dfrac{m_0}{(J+1/2)} \\ p_z(J) = p_z \dfrac{\gamma_J}{\gamma_0} \dfrac{1}{(J+1/2)} \\ p_\pm(J) = p_z \dfrac{\gamma_J}{\gamma_0} \dfrac{1}{(J+1/2)} \\ E(J) = E_0 \dfrac{\gamma_J}{\gamma_0} \dfrac{1}{(J+1/2)} \\ E(J) + m(J) = \dfrac{(\gamma_J + 1)}{(J+1/2)} m_0 \end{cases}, \tag{21}$$

where $\gamma_J$ denotes the Lorentz factor of the fermion with energy $E(J)$. Using Eq. (21) and considering that the final state is $\psi_0$, we arrive at the explicit form of matrix $\Lambda_J$ (for positive energy spinors):

$$\Lambda_J(E(J) > 0) = \begin{pmatrix} \mathbb{1} & \mathbb{0} \\ \mathbb{0} & \dfrac{\gamma_J(\gamma_0 + 1)}{\gamma_0(\gamma_J + 1)} \mathbb{1} \end{pmatrix}. \tag{22}$$

For negative energy spinors, instead, are the first two diagonal elements to be multiplied times the function $f(J) = \gamma_J(\gamma_0 + 1)/\gamma_0(\gamma_J + 1)$. Considering Eq. (9), the explicit form of function $f(J)$ corresponding to the maximum probability of bradyon-tachyon field interaction is:

$$f(J) = \frac{\gamma_J(\gamma_0 + 1)}{\gamma_0(\gamma_J + 1)} = \frac{2J + 3 + 2\sqrt{2J + 1}}{4(J + 3/2)}. \tag{23}$$

At the limit $J \to \infty$, this function tends to $1/2$. This result may be generalized for a decay from $(J + 1)$ to $J$:

$$f(J) = \frac{\gamma_{J+1}(\gamma_J + 1)}{\gamma_J(\gamma_{J+1} + 1)}. \tag{24}$$

## 5. Decay Spectrum

Supposing that the decay involves three bodies, by analogy with beta decay, we can predict the disintegration spectrum of a composite particle. Denoting the imaginary masses of the antitachyon and tachyon by $\mu$ and $\mu'$, the disintegration energy is:

$$E_\mu + E_{\mu'} = W(J) = \sqrt{p_\mu^2 c^2 - \mu^2 c^4} + \sqrt{p_{\mu'}^2 c^2 - {\mu'}^2 c^4}. \tag{25}$$

The number of states in which the negative energy tachyon has momentum within the range $[p_\mu, p_\mu + dp_\mu]$ is given by:

$$d^2N = dn_\mu dn_{\mu\prime} = \left(\frac{4\pi\Omega}{(2\pi\hbar)^3}p_\mu^2 dp_\mu\right)\left(\frac{4\pi\Omega}{(2\pi\hbar)^3}p_{\mu\prime}^2 dp_{\mu\prime}\right), \quad (26)$$

where $\Omega$ is the volume of normalization. For a given value of $E_\mu$ from Eq. (25), we get the impulse $p_{\mu\prime}$ that once differentiated gives:

$$p_{\mu\prime}^2 dp_{\mu\prime} = \frac{1}{c^3}\sqrt{(W-E_\mu)^2 - \mu\prime^2 c^4}(W-E_\mu)dW. \quad (27)$$

Substituting Eq. (27) in Eq. (26), we obtain the density of antitachyon final states when disintegration energy is within the range $[W, W + dW]$:

$$\frac{d^2N}{dWdp_\mu} = \frac{16\pi^2\Omega^2}{c^3(2\pi\hbar)^6}\sqrt{(W-E_\mu)^2 - \mu\prime^2 c^4}(W-E_\mu)p_\mu^2 dp_\mu, \quad (28)$$

Since the decay occurs at a given disintegration energy, the density of antitachyon final states may be simplified as follows:

$$\frac{dN}{dp_\mu} = \frac{16\pi^2\Omega^2}{c^3(2\pi\hbar)^6}\sqrt{(W-E_\mu)^2 - \mu\prime^2 c^4}(W-E_\mu)p_\mu^2. \quad (29)$$

Formulating the impulse as a function of energy and considering Eq. (18), we arrive at the final result which gives the decay spectrum, in terms of density of states, as a function of the antitachyon energy:

$$\frac{dN}{dE_\mu} = \frac{16\pi^2\Omega^2}{c^3(2\pi\hbar)^6}E_\mu(W-E_\mu)\sqrt{E_\mu^2 - \mu^2 c^4}\sqrt{(W-E_\mu)^2 - \left(|\mu| - \frac{\sqrt{(J+\frac{1}{2})-1}}{(J+\frac{1}{2})}m_0\right)^2 c^4}. \quad (30)$$

The first four terms are analogous to those forming the beta decay spectrum, modulated by a radical function that depends on the quantum number $J$. The spectrum given by Eq. (30) tends towards that of beta decay as the imaginary mass of tachyon tends to zero.

## 6. Majorana Infinite Component Lagrangian

The Majorana Lagrangian density can be thought of as an infinite sum of Dirac Lagrangians of particles with decreasing mass [15]:

$$\mathcal{L} = \sum_J \left[ \Lambda_J^{-1} \bar{\psi}_0 \left( i\gamma^\mu \partial_\mu - \frac{m_0}{(J+1/2)} \right) \Lambda_J^{-1} \psi_0 \right]. \tag{31}$$

Considering that $\Lambda_J$ is a diagonal matrix, the following equation holds:

$$\gamma_J^\mu = \left(\Lambda_J^{-1}\right)^t \gamma^\mu \Lambda_J^{-1} = \Lambda_J^{-1} \gamma^\mu \Lambda_J^{-1} = f^{-1}(J) \gamma^\mu \tag{32}$$

and Lagrangian (25) becomes:

$$\mathcal{L} = \sum_J f^{-1}(J) \left[ \bar{\psi}_0 (i\gamma^\mu \partial_\mu - m_0) \psi_0 + \frac{(J-1/2)}{(J+1/2)} m_0 \bar{\psi}_0 \psi_0 \right]. \tag{33}$$

This Lagrangian reduces to that of Dirac for half-integer spin, while, for increasing $J$, each component of the infinite sum represents the Lagrangian of interaction between the Dirac field of the initial particle and the tachyon field. Therefore, for each component, the interaction term is implicitly contained in the Lagrangian through the function $f^{-1}(J)$ and the coefficient $(J-1/2)/(J+1/2)$.

Since we supposed that the decay of the composite particle with mass $m(J+1)$ is mediated by a vector boson and that the mechanism is similar to that of beta decay, neglecting the emission of the Goldstone luxon, the Lagrangian can be written as:

$$\mathcal{L}(J+1 \to J) = K(J+1)\{N_\mu [(\bar{\psi}_J \gamma_{J+1}^\mu \psi_{J+1}) + (\bar{\psi}_t \gamma_{J+1}^\mu \psi_{t'})] + h.c.\}, \tag{34}$$

where $K(J+1)$ is the coupling constant, $N_\mu$ is the $SU(2)$ matrix representing the mediator boson, and $\psi_t$ and $\psi_{t'}$ are the tachyon fields with imaginary masses satisfying the constraint (18). Due to the lack of experimental data concerning the existence of tachyons, the term $(\mathbb{1} \pm \gamma_5)$, which determines whether they are of the right-handed or left-handed type, has been omitted. Since the proposed decay mechanism, in the current state of things, is only a hypothesis for a more in-depth study, in this work, one decides to not go further in order to avoid entering into a too speculative ambit.

## 6. Particle Mass Oscillation

Neutrino flavour oscillation [27-28] can be faced by the two-state system method, where the particle occupies two pure mass states that overlap and whose eigenvalues are described by the mixing matrix [29]. The system, as a whole, is described by the superposition of the energy states, and its *transmutation* occurs only after the introduction of a perturbation $W$. In our case, this perturbation is represented by the interaction with the negative energy tachyon field:

$$\omega = \frac{\sqrt{4|W_{12}|^2 + (E_1 - E_2)^2}}{2\hbar}, \tag{35}$$

where $E_1$ is the energy of the initial state and $E_2$ that of the final state. Replacing $\omega$ with the angular resonance frequency given by Eq. (11), and considering that:

$$\begin{cases} E_1 = E_J = \gamma_J m_0 c^2 = \dfrac{(J + 1/2)}{\hbar\sqrt{2J}} m_0 c^2 \\ E_2 = E_{J+1} = \gamma_{J+1} m_0 c^2 = \dfrac{(J + 3/2)}{\hbar\sqrt{2(J + 1)}} m_0 c^2 \end{cases}, \tag{36}$$

from Eq. (35), we get:

$$|W_{J,J+1}| = \sqrt{\frac{(J + 1/2)^2}{2J} - \left[\frac{(J + 1/2)}{4\sqrt{2J}} - \frac{(J + 3/2)}{4\sqrt{2(J + 1)}}\right]^2} m_0 c^2. \tag{37}$$

Eq. (37) gives the minimum perturbation needed for the mass oscillation process to take place and can be read as the interaction energy between the particle with mass $m_J$ and the tachyon field. Since $m_J = m_0 f(J)/(J + 1/2)$, the term representing the mass oscillation is:

$$\Delta m(J \to J + 1) = -m_0 \frac{(J + 1/2)(J + 5)(J + 3/2 + \sqrt{2J + 2})}{j + 5/2 + \sqrt{2J + 4}}. \tag{38}$$

The probability that the mass oscillation occurs is given by [30]:

$$P(J \to J + 1) = \sin^2(2\theta)\sin^2\left(\frac{\Delta m^2 c^3}{4\hbar E}x\right) = \sin^2(2\theta)\sin^2(\omega_{rest}t) \tag{39}$$

from which we get the angular resonance frequency in terms of mass oscillation:

$$\omega_{res} = \frac{\Delta m^2 c^3}{4\hbar E}\frac{x}{t} = \frac{\Delta m^2 c^3}{4\hbar E} u_{max}, \qquad (40)$$

where $u_{max}$ is particle velocity at which the probability on *transmutation* is maximum. Substituting the explicit form of $\omega_{res}$ in Eq. (40) and considering that:

$$u_{max} = \frac{(J - 1/2)}{(J + 1/2)} c, \qquad (41)$$

we arrive at the following result:

$$\Delta m(J \to J + 1) = -m_0 2\left(J + \frac{1}{2}\right)\sqrt{\frac{\left(J + \frac{1}{2}\right)}{2\left(J - \frac{1}{2}\right)}}. \qquad (42)$$

Eq. (42) is completely different compared to Eq. (38), and this proves that mass oscillation is not only the result of mixing between the two pure mass states but that the particle also undergoes decay. Therefore, our theory predicts charge-parity (CP) violation through mixing-decay interference [31], a result that is in agreement with experimental data. Plotting the difference between Eq. (42) and Eq. (38) vs quantum number $J$, we obtain the contribution due to the decay mechanism compared to that due to mixing:

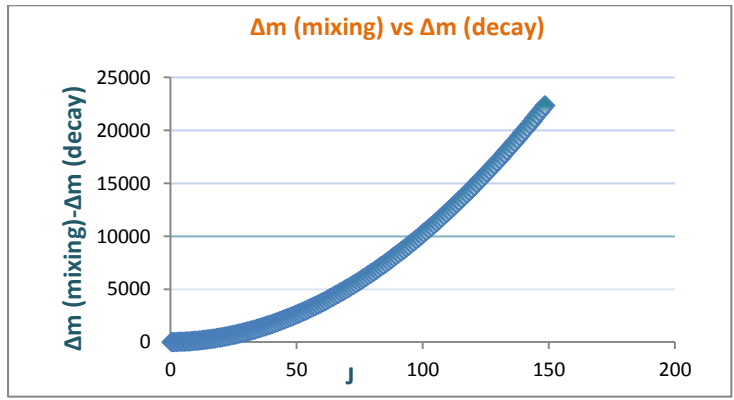

**Figure 1**: Mixing and decay contributions to the disintegration mechanism.

From Fig. 1, we see that, for high values of quantum number $J$, the contribution due to decay overcomes that of mixing, and they are comparable only for a small value of $J$. Based on these results, we can state that decay is the main cause of particle mass oscillation.

## 7. Discussion

The formulation of a theory of interaction between ordinary and tachyonic matter, based on the infinite-dimensional representation of the Lorentz group [32], leads to equations capable of describing the mass oscillation of the particles. Starting from Majorana equation, revisited in the picture of quantum field theory, we have proved that a decreasing spectrum of real masses, associated to composite particles formed by a bradyonic fermion and a set of tachyon-antitachyon pairs, can be obtained. These new entities are stable only at very high energy conditions and tend to disintegrate quickly as their velocity decreases. The proposed mechanism of disintegration is very similar to β-decay and could be the key to explaining the mass oscillation of light particles, such as the neutrino. In fact, the neutrino's very small mass and its velocity, which is very close to the speed of light, represent the optimal conditions of interaction with the tachyon field. Therefore, our theory predicts a high probability of mass oscillation for the neutrino, a phenomenon that experience has proved to exist [27-28]. This result suggests that the study of neutrino flavour oscillation is the best place where massive tachyons could be experimentally observed. This is the reason why we also investigated dynamic decay to obtain the emission spectrum of a tachyon with negative energy, giving to experimentalists a theoretical tool with which to prove the correctness of such a mechanism of mass oscillation.

Thinking to the Majorana field as the sum of the Dirac field and the infinite half-integer tachyon fields with increasing imaginary mass, we arrive at a very simple Lagrangian density that implicitly contains the interaction term. In this scenario, the mechanism of *transmutation* of a particle with a given $J$ into a higher or lower $J$ can be thought of as being mediated by a boson formed by a tachyon and an antitachyon whose total mass must comply with constraint given by Eq. (3). Everything happens as if the field of mass $m(J)$ interacts with the field of mass $m(J \pm 1)$ through a tachyonic *pion*, in full analogy with the mechanism of the strong force that binds protons and neutrons.

## 8. Conclusion

This study has been developed by introducing speculative ideas that refer to physical phenomena that have not yet been confirmed by experimental physics or for which there is no scientific project dedicated to them. The results obtained, however, have many points in common with those predicted by the Standard Model and show that theoretical physics can and must go beyond current theories to explain experimental phenomena that still lack a rational explanation. In this scenario, the tachyon matter can be just the element needed to extend the range of action of the Standard Model, so as to explain unresolved problems such as the mass oscillation of the particles. In this regard, in the present work, we have focused attention on the case of the neutrino, but attention can also be given to the imperceptible mass variations observed for elementary particles belonging to the families of baryons and mesons [34-35]. It is precisely from the study of these phenomena, often difficult to observe or that can be hidden by instrumental errors, that we could indirectly and effectively verify the existence of tachyon matter. The interaction of tachyons with ordinary matter is certainly of a very weak nature [36], and this work is a further proof, but additional research through the study of mass particle oscillation is our proposal to discover its nature and, thus, open the door for a wider research that could also involve dark matter [37].